\documentstyle[prl,preprint,aps,epsfig]{revtex}
\begin{document}
\draft
\title{Isospin and Z$^{1/3}$ Dependence of the Nuclear Charge Radii}
\author{S.Q. Zhang$^{1}$, J. Meng$^{1,2,3}$\thanks{e-mail: mengj@pku.edu.cn},
S.-G. Zhou$^{1,2,3}$, and J.Y. Zeng$^{2,3,4}$}
\address{${}^{1}$Department of Technical Physics, Peking University, Beijing 100871}
\address{${}^{2}$Institute of Theoretical Physics, Chinese Academy of Sciences, Beijing 100080}
\address{${}^{3}$Center of Theoretical Nuclear Physics, National Laboratory of \\
       Heavy Ion Accelerator, Lanzhou 730000}
\address{${}^{4}$Department of Physics, Peking University, Beijing 100871\\ }
\date{\today}
\maketitle

\begin{abstract}
Based on the systematic investigation of the data available for $A \geq 40$,  a
$Z^{1/3}$ dependence for the nuclear charge radii is shown to be superior to the
generally accepted $A^{1/3}$ law. A delicate scattering of data around $R_c/Z^{1/3}$
is infered as owing to the isospin effect and a linear dependence of $R_c/Z^{1/3}$ on
$N/Z$ ( or $(N-Z)/2$ ) is found. This inference is well supported by the microscopic
Relativistic Continuum Hartree-Bogoliubov (RCHB) calculation conducted for the proton
magic Ca, Ni, Zr, Sn and Pb isotopes including the exotic nuclei close to the neutron
drip line.  With the linear isospin dependence provided by the data and RCHB theory, a
new isospin dependent $Z^{1/3}$ formula for the nuclear charge radii is proposed.
\end{abstract}


Nuclear radius is one of the most fundamental bulk properties of an atomic
nucleus\cite{BM69,RS80}. Among all the size quantities describing nucleus, nuclear
charge radii has been investigated by various techniques and methods experimentally
\cite{FBH95,AHS87,OT89,BC95,SAE99,LB99,VJV87}, including the
 muonic atom spectroscopy\cite{FBH95}, isotope shift of
 optical and K X-ray spectroscopy\cite{AHS87,OT89,BC95,SAE99,LB99}
 and high energy elastic electron scattering \cite{VJV87},
etc.. Recently more and more nuclei far from the $\beta$-stability line become
accessible experimentally thanks to the development of radioactive ion beam
facilities\cite{Ta.95,HJJ.95}. The nuclear size connected with exotic phenomena such
as skin and halo have become a hot topic. The understanding of its property has
importance not only in nuclear physics, but also in other fields such as astrophysics
and atomic physics, etc. With its accuracy, the study of the nuclear charge radii is
very important to understand not only the proton distribution inside the nucleus but
also the halo and skin. Particularly if one can get a simple and reliable formula for
nuclear charge radii, it will be very useful to extract the de-coupling of proton and
neutron in the exotic nuclei and provide information for the effective nucleon-nucleon
interaction widely used in all the nuclear models. Here in this letter the available
experimental charge radii data for $A \geq 40$ will be examined and its global
behavior will be studied. Instead of the widely accepted $A^{1/3}$ law, a new
$Z^{1/3}$ formula with isospin effect will be proposed.

Based on the consideration of the nuclear saturation property,
nuclear charge radii $R_c$ are usually described by the $A^{1/3}$ law \cite{BM69,RS80}
\begin{equation}
R_c=r_A A^{1/3},
\label{eqn.A}
\end{equation}
where $A$ is the mass number and $R_c=\sqrt{\frac{5}{3}}\ \langle r^2 \rangle ^{1/2}$,
with $\langle r^2 \rangle ^{1/2}$ the root-mean square (rms) charge radius. For very
light nuclei, because of their small $A$ and large fluctuation in charge distribution
due to the shell effect with short period, it seems that the charge distribution
radius as a bulk property has little meaning. A detail analysis of charge radius data
for $A\geq 40$ shows that $r_A$ is by no means a constant, but systematically
decreases with $A$; i.e., $r_A \approx 1.31$ fm for light nuclei $(A\sim 40)$ and
$r_A\approx 1.20$ fm for very heavy nuclei (see upper left panel in Fig. 1). This fact
implies that some physics is missing in Eq. (\ref{eqn.A}).

A definite evidence of the violation of $A^{1/3}$ law is also found in
the measurements of isotope shift in mean square charge radii
 \cite{ZE57,PP93}.
In particular, $\delta\langle r^2\rangle_{A+2,A}$ values (associated with an
addition of two neutrons) are often found to be considerably smaller
compared to what is expected from the $A^{1/3}$ law
($\delta \langle r^2\rangle_{A+2,A}=\frac{4}{3A}\langle r^2\rangle_{A}$).
A typical example is that the observed charge radii of the calcium isotopes
$^{40-50}$Ca remain almost the same (except a very little change induced by
deformation or shell effect), though the mass number $A$ has changed significantly.
In contrast, there is also evidence that the observed
$\delta\langle r^2\rangle_{A+2,A}$ values (associated with the addition of two
protons) are often greater than what is expected from the $A^{1/3}$ law
(e.g., $\delta\langle r^2\rangle$ for $^{46}$Ti$-^{44}$Ca, $^{50}$Ti$-^{48}$Ca, etc.).

Along the $\beta$-stability line, the ratio $Z/A$ gradually decreases with $A$, i.e.,
for light nuclei $Z/A\approx 1/2$, and for the heaviest $\beta$-stable nucleus
$^{238}_{\ 92}$U, $(Z/A)^{1/3}\approx 0.7285$, thus $(1/2)^{1/3}/(Z/A)^{1/3}\approx
1.09$, which is very close to the $r_A$ ratio $1.30/1.20$ shown in upper left panel of
Fig. 1. A naive point of view is that the charge radius of a nucleus may be more
directly related to its charge number $Z$, rather than its mass number $A$. Therefore,
compared to the $A^{1/3}$ law, a $Z^{1/3}$ dependence for nuclear charge radii may be
more reasonable
\begin{equation}
R_c = r_Z Z^{1/3}.
\label{eqn.Z}
\end{equation}
as noted in Ref. \cite{ZE57}. An analysis of the very
limited data of charge radii then available
showed that $r_Z$ remains almost a constant, i.e., $r_Z=1.65(2)$ fm for $A\geq 40$.
The $Z^{1/3}$ dependence of nuclear charge radii was also used
to modify the Coulomb energy term
in the semi-empirical nuclear mass formula \cite{TS80}, and it was found
that the agreement between
the calculated and experimental results was improved.
Moreover, the $A^{-1/3}$ law for the nuclear giant ( monopole,
dipole and quadrupole ) resonance energy ($\propto 1/R$)
also could be improved, if the $A^{-1/3}$ dependence
is replaced by a $Z^{-1/3}$ dependence \cite{ZE82}.

In the past two decades, a vast amount of new experimental information on the
electromagnetic structure of nuclear ground states of many nuclei has become available
\cite{FBH95,AHS87,OT89,BC95,SAE99,LB99,VJV87}, and accuracy has been improved. In
particular the muon factories at Los Alamos (LAMPF) and at Villigen (PSI, formerly
SIN) started their operation at 1974. Almost all stable
 nuclei have been measured by the muonic X-ray transition
 technique and the corresponding charge radii have been
 rather accurately deduced (the experimental relative error is
 about $10^{-3}$). Moreover, modern techniques for optical
 isotope shift measurements have made it possible to reach
 even short-lived (down to 1 s) unstable isotopes\cite{FBH95}.
Therefore, it is worthwhile to reexamine the fundamental property of nuclei and to
investigate whether the vast amount of improved experimental results follow the
$Z^{1/3}$ dependence. The values of measured $\langle r^2\rangle^{1/2}$ for 536 nuclei
with $A\geq 40$ compiled in
 Ref.\cite{FBH95,AHS87,OT89,BC95,SAE99,LB99,VJV87}
are analyzed in Fig. 1 by using the $A^{1/3}$ and $Z^{1/3}$ dependence, respectively.
The dependence of charge radii on the quadrupole deformation $\beta$ has been taken
into account for the rare-earth deformed nuclei, i.e. \cite{BM69}
\begin{equation}
r_A=r_{Ad}(1+\frac{5}{8\pi}\beta^2), \ \ \ \ \
r_Z=r_{Zd}(1+\frac{5}{8\pi}\beta^2),
\label{eqn.dc}
\end{equation}
and for spherical nuclei ($\beta=0$): $r_A=r_{Ad}$, $r_Z=r_{Zd}$, and the values of
$\beta$ are taken from Ref.\cite{VJV87,MNM95}.

In the upper left and right panels of Fig.1, the charge radii for the
 most stable 159 nuclei with $A\geq 40$ along the $\beta$-stability
 line have been
 analyzed by using the $A^{1/3}$ and $Z^{1/3}$ dependence. In the
 middle left and right panels, the same has been done for
 the measured $\langle r^2\rangle^{1/2}$ for 536 nuclei
 with $A\geq 40$. Two significant features can be observed:
 (A) On the one hand, the agreement between the data and the
 calculated results using the $Z^{1/3}$ dependence is much
 better than that
 using the $A^{1/3}$ law, i.e., while there exists a global
 regular decrease of $r_{Ad}$ with $A$, $r_{Zd}$ nearly remains
 constant ($r_{Zd}=1.631(11)$ fm). The relative rms deviations
 $\sigma$ for the $Z^{1/3}$ dependence
 ($\sigma= 7.57 \times 10^{-3}$ for stable nuclei and
 $1.00 \times 10^{-2}$ for 536 nuclei ) are much less than
 that for the $A^{1/3}$ law ($\sigma=1.90\times 10^{-2}$
 for stable nuclei and $1.63 \times 10^{-2}$ for 536 nuclei ).
 (B) On the other hand, though the rms deviation for the $Z^{1/3}$
 dependence is significantly reduced, an isospin induced scattering
 of the data in the middle panels in Fig. 1 can
 be also observed compared with that in the top panels.
 In fact, $r_{Zd}$ generally increases with $N$ for most
 isotopic chains, e.g. for $^{90-96}_{40}$Zr, $^{92-100}_{42}$Mo,
 $^{96-104}_{44}$Ru,
 $^{102-110}_{46}$Pd, $^{106-116}_{48}$Cd, $^{112-124}_{50}$Sn,
 $^{122-130}_{52}$Te, $^{124-136}_{54}$Xe, $^{142-148}_{60}$Nd,
 $^{144-154}_{62}$Sm, $^{154-160}_{64}$Gd, etc, (except for only a
 few lighter isotopic chains, e.g., $^{78-86}_{36}$Kr,
 $^{84-88}_{38}$Sr, and a small anomalous decrease of $r_{Zd}$
 with $N$ due to the shell closure at $N=50$ is observed). Therefore, it seems
 necessary to investigate an isospin dependent correction for the
 scattering of $r_{Zd}$. In Ref. \cite{PP93,WPP98}, the
 isospin effect has been considered based on the $A^{1/3}$ law.
 However, considering the fact that the $Z^{1/3}$ dependence can
 describe the nuclear charge radii much better than the
 $A^{1/3}$ law, we take the $Z^{1/3}$ dependence as a more
 reasonable starting point for describing the isospin dependence
 of nuclear charge radii.

To confirm that isospin dependent $Z^{1/3}$ formula to be developed for nuclear charge
radii mentioned above is also valid for nuclei far from the $\beta$-stability line,
the charge radii in exotic nuclei is needed. However, as no such data is available,
what we can do is to require that our new isospin dependent $Z^{1/3}$ formula should
assort with a reliable and microscopic nuclear model.

The fully self-consistent and microscopic relativistic continuum Hartree-Bogoliubov
(RCHB) theory, which is an extension of the relativistic mean field (RMF)
\cite{SW86,RE89,RI96} and the Bogoliubov transformation in the coordinate
representation\cite{ME98},  is a good candidate for the present purpose. The RCHB
theory can describe satisfactorily the ground state properties for nuclei both near
and far from the $\beta$-stability line. A remarkable success of the RCHB theory is
the self-consistent reproduction of the halo in $^{11}$Li \cite{MR96} and prediction
of the exotic phenomenon - giant halo \cite{MR98}. In combination with the Glauber
model, the RCHB theory successfully reproduces the interaction cross section in Na
isotopes \cite{MTY.97} and the charge changing cross section of C, N, O, F isotopes
(ranging from the $\beta$-stability line to the neutron drip line) on the target of
$^{12}$C at 930 MeV/u \cite{Chu00,ME00}. These successes encourage us to apply the
RCHB theory for the description of charge radii of nuclei both close to and far from
the $\beta$-stability line and check its validity for the data available and provide
information for nuclei far away from the stability line.

The detailed formalism and numerical techniques of the RCHB theory can
be found in Ref.\cite{ME98} and the references therein. In the present
calculations, we follow the procedures in Ref.\cite{ME98,MR98,MTY.97}
and solve the RCHB
equations in a box with the size $R=20$ fm and a step size of 0.1 fm.
The parameter set NL-SH \cite{SNR93} is used, which aims at describing both the
stable and exotic nuclei. The density dependent $\delta$-force in the pairing channel
with $\rho_0=0.152$ fm$^{-3}$ is used and its strength $V_0$
is fixed by the Gogny force as in Ref.\cite{ME98}. The contribution from continua is
restricted within a cut-off energy $E_{cut}\sim 120$MeV.


As typical examples, we studied the even-even Ca, Ni, Zr, Sn and Pb isotopes ranging
from the $\beta$-stability line to neutron drip line.
 The two neutron separation energies $S_{2n}$ is one of the
 essential quantities to test a nuclear model. In
 Fig. 2, the calculated $S_{2n}$ (open symbols) of the
 even-even Ca, Ni, Zr, Sn and Pb isotopes by the RCHB theory are
 compared with the data available (solid symbols) \cite{Audi95},
 where a satisfactory agreement is seen. Particularly
 the deviation between the calculated binding energies with
 the data available is within $1\%$. In the present calculation,
 the neutron drip-line nuclei are predicted at $^{72}$Ca, $^{98}$Ni,
 $^{140}$Zr and $^{176}$Sn, respectively. In the $S_{2n}$ {\it versus} $N$ curve for each isotopic chain, there
are some kinks due to the neutron shell or subshell closure. For example, the closed
shells at $N=$ 20, 28 and subshell at $N=40$ correspond to kinks in the $S_{2n}$ {\it
versus} $N$ curve for Ca isotopes at $^{40}$Ca, $^{48}$Ca and $^{60}$Ca, respectively.
While the kink at $N=20$ for $^{40}$Ca may be also due to the Wigner term for
$N=Z=20$. However, there are no kinks at $^{70}$Ca and $^{176}$Sn,
which indicate the
disappearance of magic number $N=50$ and $126$ for these nuclei
in RCHB.

The rms charge radii $\langle r^2\rangle^{1/2}$ obtained from the RCHB theory (open
symbols) and the data available (solid symbols) for the even-even Ca, Ni, Zr, Sn and
Pb isotopes are given in Fig. 3. As it could be seen, the RCHB calculations reproduce
the data very well ( within 1.5$\%$). For a given isotopic chain, an approximate
linear $N$ dependence of the calculated rms charge radii $\langle r^2\rangle^{1/2}$ is
clearly seen in Fig. 3, which shows that the variation of $\langle r^2\rangle^{1/2}$
for a given isotopic chain deviates from both the the simple $Z^{1/3}$ dependence and
the simple $A^{1/3}$ law (denoted by dashed lines in Fig. 3). Therefore, a strong
isospin dependence of nuclear charge radii is necessary for nuclei with extreme $N/Z$
ratio.

In Fig. 4, the experimental and RCHB predicted $r_{Zd} = R_c / Z^{1/3}$ for the proton
magic isotopes are presented as a function of $\eta=N/Z$. It is clearly seen that the
coefficient $r_{Zd}$ increases linearly with $\eta$ (except some deviations due to
deformation or shell effect) and the slopes are nearly the same for these isotopic
chains. The linear $\eta$ ( or isospin $T_Z=(N-Z)/2$ ) dependence of $r_{Zd}$ for an
isotopic chain may be understood as the effect of the first order perturbation
correction of nuclear wave function due to an isospin $T_Z$ dependent interaction
\cite{WI57}. Based on the analysis of data in the middle and upper panels
 of Fig. 1 and RCHB prediction in Fig. 3 and 4, we propose the following
 isospin dependent $Z^{1/3}$
formula for nuclear charge radii:
\begin{equation}
R_c=a\,Z^{1/3}\left[1+b(\eta-\eta^*)\right], \ \ \ \
\eta=N/Z,
\label{eqn.rc}
\end{equation}
where $\eta^*$ is $\eta=N/Z$ for the nuclei along the $\beta$-stability line which can
be directly extracted from the nuclear mass formula \cite{BM69},
$a=r^{*}_{Zd}(1+\frac{5}{8\pi}\beta^2)$, $r^{*}_{Zd}= 1.631(11)$ fm
 obtained in upper right panel of Fig.1,
 and $b=0.062(9)$ obtained from the least square fitting.

The analysis of the available data using Eq. (\ref{eqn.rc})
 with $r_{Zd}^{*}$ and $b$ thus obtained is displayed in the
 lower right panel of Fig. 1. It is found the data are
 reproduced better by Eq. (\ref{eqn.rc}) than by Eq. (\ref{eqn.Z})
 (the rms deviation is reduced by about $40\%$).
 The same has been done for $A^{1/3}$ dependence but with less
 success ( see the lower left panel of Fig. 1 ).
 In Ref. \cite{PP93,WPP98}, a similar equation for
 $A^{1/3}$ dependence has been used to describe the isospin
 dependence of charge radii and a better agreement has been
 achieved by fitting both $r_A$ and $b$ simultaneously. But
 then a simple explanation for $r_A$ (saturation property) and
 $b$(isospin effect) in  Eq. (\ref{eqn.rc}) is missing.
 It is expected that the modified $Z^{1/3}$ formula (Eq. (\ref{eqn.rc})) will
become more useful with more and more data obtained for the nuclei far from the
$\beta$-stability line.

In summary, we have systematically investigated the nuclear charge radii with $A \geq
40$. It is clearly seen that the $Z^{1/3}$ dependence is superior to the $A^{1/3}$
law. A delicate scattering of data around $R_c/Z^{1/3}=1.631$ is infered as owing to
the isospin effect and a linear dependence of $R_c/Z^{1/3}$ on $N/Z$ ( or $(N-Z)/2$ )
is found. This inference are well supported by the microscopic RCHB calculation
conducted for the proton magic Ca, Ni, Zr, Sn and Pb isotopes including the exotic
nuclei close to the neutron drip line, which reproduce $S_{2n}$ and nuclear charge
radii data available well. With the linear dependence of the coefficient $r_{Zd}$ on
$N/Z$ (or $(N-Z)/2$ ) read from data and RCHB theory, a new isospin dependent
$Z^{1/3}$ formula for the nuclear charge radii is proposed, which improves the
description of the data available for nuclei near the $\beta$-stability line and could
be very useful for new data obtained for nuclei far from the $\beta$-stability line.


We would like to thank A.Arima, B.A.Brown, I.Hamamoto,
K.Matsuyanagi, W.Q.Shen, I.Tanihata, B.Tsang and V.Zelevinsky
for delightful discussions. This work was  partly supported by
the Major State Basic Research
Development Program Under Contract Number G2000077407 and the
the National Natural Science Foundation of China under Grant
Nos. 10025522, 19847002 and 19935030.


\begin{figure}
\centerline{\epsfig{figure=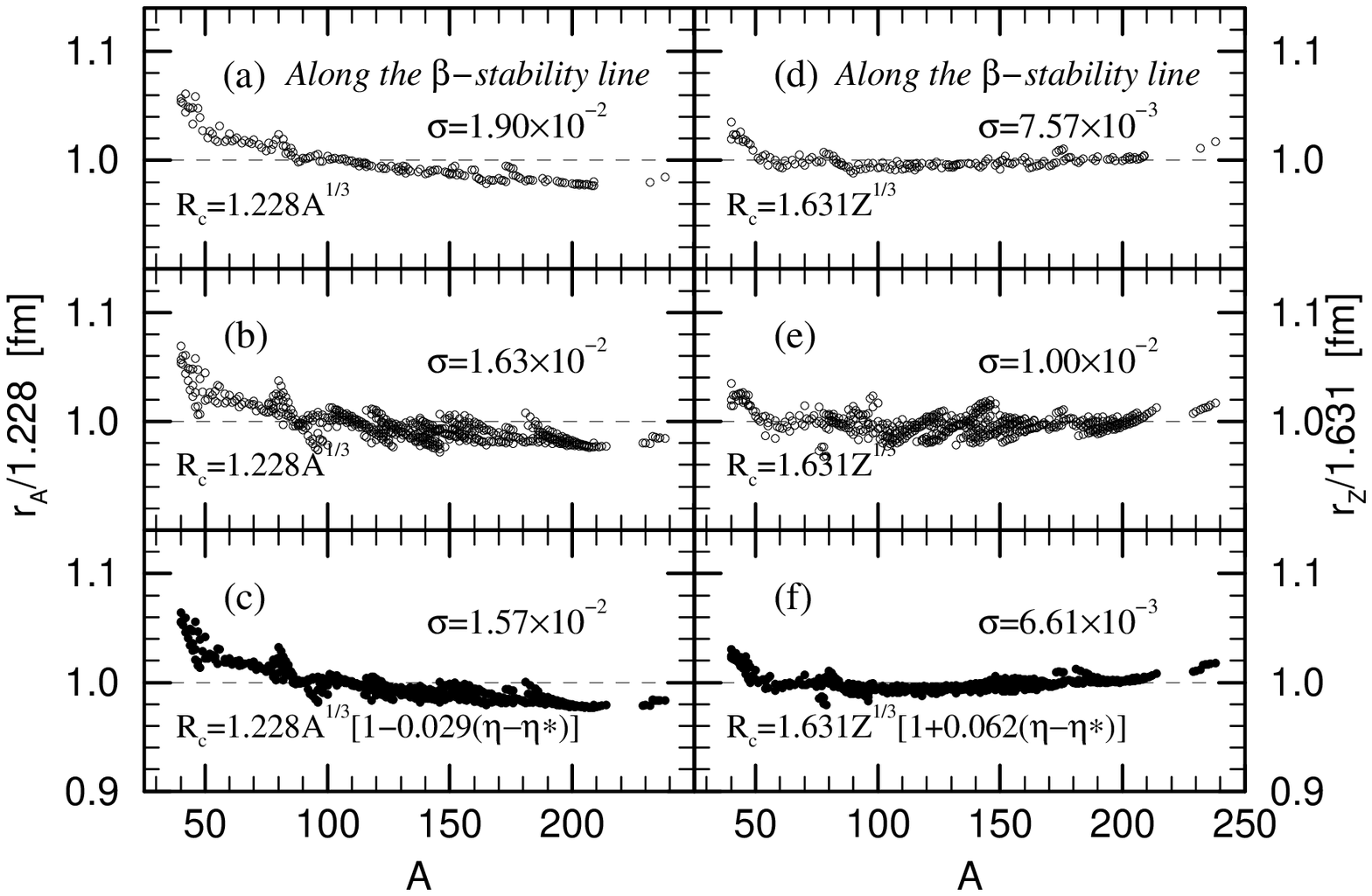,width=10.cm}} \caption{The
 nuclear charge radius
 data for $r_{A}$ in $A^{1/3}$ and $r_{Z}$ in $Z^{1/3}$ law
 with and without isospin dependence, for the details see the text.
 } \label{fig.expr}
\end{figure}

\begin{figure}
\centerline{\epsfig{figure=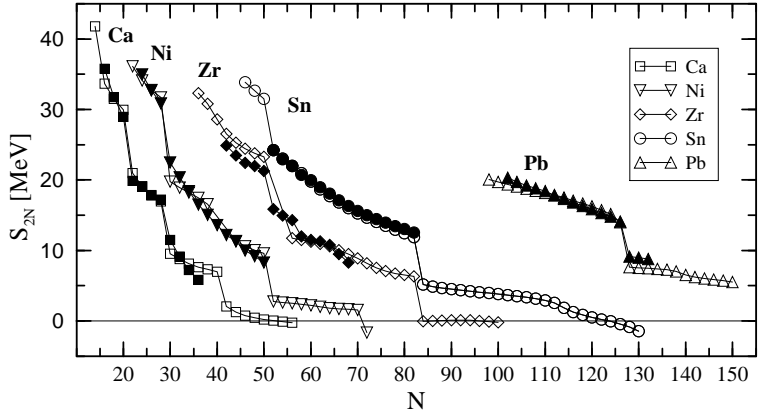,width=10.0cm}} \caption {Two-neutron separation
energies $S_{2n}$ of even Ca, Ni, Zr, Sn, Pb isotopes as a function of $N$, including
the data (solid symbols) from Ref.[28] and the RCHB calculation with a $\delta$-force
(open symbols).} \label{fig.s2n}
\end{figure}

\begin{figure}
\centerline{\epsfig{figure=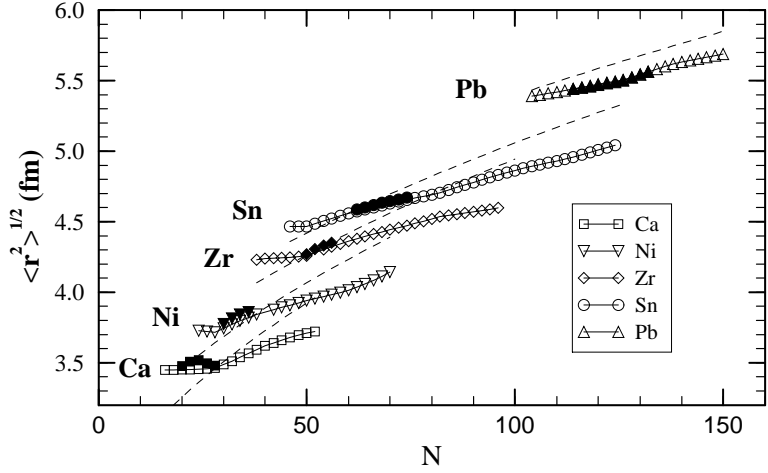,width=10.0cm}} \caption{The rms charge radii {\it
versus} the neutron number $N$ for even-even Ca, Ni, Zr, Sn, Pb isotopes. The RCHB
calculation with $\delta$-force is represented by open symbols, while the
corresponding data is denoted by solid symbols.
 The dashed lines represent the
predictions by the $A^{1/3}$ law with $r_{Ad}=1.228$ fm.}
 \label{fig.rc}
\end{figure}

\begin{figure}
\centerline{\epsfig{figure=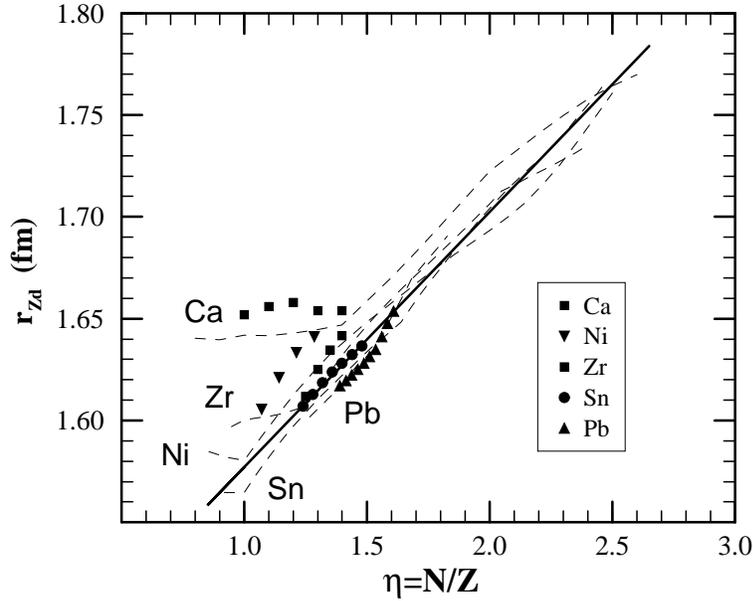,width=10.cm}}
\caption{The experimental (solid symbols) and RCHB predicted ( dashed lines )
 coefficient $r_{Zd}=R_c/Z^{1/3}$ for the nuclear
 charge radii as a function of isospin quantity $\eta=N/Z$
 in even-even Ca, Ni, Zr, Sn and Pb isotopes.
 An asymptotic behavior is drawn as a solid line.}
\label{fig.rp}
\end{figure}

\end{document}